\documentclass[aps,prc,floatfix,nofootinbib,showpacs,twocolumn,superscriptaddress]{revtex4-1}

\usepackage{amsmath}
\usepackage{amssymb}
\usepackage{graphicx}
\usepackage{color}
\usepackage{xspace}

\usepackage[mathscr,scaled=1.15]{urwchancal}
\DeclareFontFamily{OT1}{pzc}{}
\DeclareFontShape{OT1}{pzc}{m}{it}%
{<-> s * [1.15] pzcmi7t}{}
\DeclareMathAlphabet{\mathpzc}{OT1}{pzc}{m}{it}

\definecolor{purple}{rgb}{0.5,0,0.5}
\definecolor{blue}{rgb}{0.0,0,0.9}
\definecolor{prdblue}{rgb}{0.133,0.118,0.498}
\usepackage[colorlinks=true, pdfstartview=FitV, linkcolor=prdblue, citecolor= prdblue, urlcolor=prdblue]{hyperref}

\begin{document}

\title{Off-shell persistence of composite pions and kaons}

\author{Si-Xue Qin}
\affiliation{Physics Division, Argonne National Laboratory, Argonne,
Illinois 60439, USA}

\author{Chen Chen}
\affiliation{Instituto de F\'isica Te\'orica, Universidade Estadual Paulista, 01140-070 S\~ao Paulo, Brazil}

\author{C\'edric Mezrag}
\affiliation{Physics Division, Argonne National Laboratory, Argonne,
Illinois 60439, USA}

\author{Craig D.~Roberts}
\affiliation{Physics Division, Argonne National Laboratory, Argonne,
Illinois 60439, USA}

\date{08 February 2017}

\begin{abstract}
In order for a Sullivan-like process to provide reliable access to a meson target as $t$ becomes spacelike, the pole associated with that meson should remain the dominant feature of the quark-antiquark scattering matrix and the wave function describing the related correlation must evolve slowly and smoothly.  Using continuum methods for the strong-interaction bound-state problem, we explore and delineate the circumstances under which these conditions are satisfied: for the pion, this requires $-t \lesssim 0.6\,$GeV$^2$, whereas $-t\lesssim 0.9\,$GeV$^2$ will suffice for the kaon.  These results should prove useful in planning and evaluating the potential of numerous experiments at existing and proposed facilities.
\end{abstract}

\maketitle

\begin{figure}[t]
\vspace*{2ex}
\begin{center}
\includegraphics[width=0.55\linewidth]{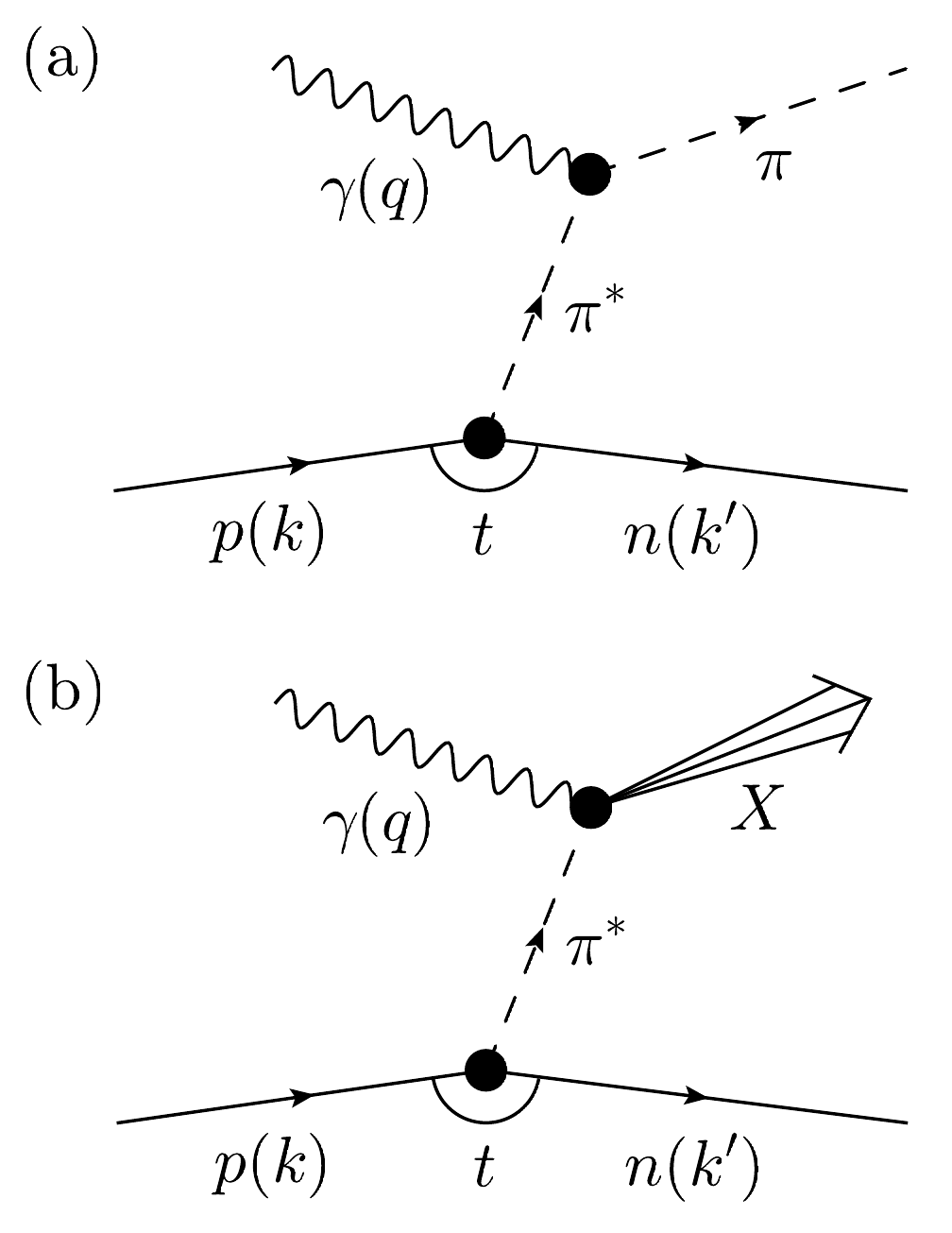}
\end{center}
\caption{Sullivan processes, in which a nucleon's pion cloud is used to provide access to the pion's (a) elastic form factor and (b) parton distribution functions.  $t=-(k-k^\prime)^2$ is a Mandelstam variable and the intermediate pion, $\pi^\ast(P=k-k^\prime)$, $P^2=-t$, is off-shell.
\label{figSullivan}}
\end{figure}

\noindent\textbf{1.$\;$Introduction}.
%
The notion that a nucleon possesses a meson cloud is not new \cite{Fermi:1947zz}.  In effect, this feature is kindred to the dressing of an electron by virtual photons in quantum electrodynamics \cite{Feynman:1949zx} or the existence of dressed quarks with a running mass generated by a cloud of gluons in quantum chromodynamics (QCD) \cite{Lane:1974he, Politzer:1976tv, Bhagwat:2003vw, Bowman:2005vx, Bhagwat:2006tu}.  Naturally, any statement that each nucleon is accompanied by a meson cloud is only meaningful if observable consequences can be derived therefrom.  A first such suggestion is canvassed in Ref.\,\cite{Sullivan:1971kd}, which indicates, \emph{e.g}.\ that a calculable fraction of the nucleon's anti-quark distribution is generated by its meson cloud.  Mirroring this effect, one may argue that a nucleon's meson cloud can be exploited as a target and thus, for instance, the so-called Sullivan processes can provide a means by which to gain access to the pion's elastic electromagnetic form factor \cite{Volmer:2000ek, Horn:2006tm, Tadevosyan:2007yd, Horn:2007ug, Blok:2008jy}, Fig.\,\ref{figSullivan}(a), and also its valence-quark parton distribution functions (PDFs) \cite{Holt:2000cv, Holt:2010vj, Keppel:2015}, Fig.\,\ref{figSullivan}(b).

One  issue in using the Sullivan process as a tool for accessing a ``pion target'' is that the mesons in a nucleon's cloud are virtual (off-shell) particles.  This concept is readily understood when such particles are elementary fields, \emph{e.g}.\ photons, quarks, gluons.  However, providing a unique definition of an off-shell bound-state in quantum field theory is problematic.

Physically, for both form factor and PDF extractions, $t<0$ in Figs.\,\ref{figSullivan}, so the total momentum of the $\pi^\ast$ is spacelike.\footnote{We use a Euclidean metric:  $\{\gamma_\mu,\gamma_\nu\} = 2\delta_{\mu\nu}$; $\gamma_5= \gamma_4\gamma_1\gamma_2\gamma_3$, tr$[\gamma_5\gamma_\mu\gamma_\nu\gamma_\rho\gamma_\sigma]=-4 \epsilon_{\mu\nu\rho\sigma}$; $\sigma_{\mu\nu}=(i/2)[\gamma_\mu,\gamma_\nu]$; $a \cdot b = \sum_{i=1}^4 a_i b_i$; and $P_\mu$ spacelike $\Rightarrow$ $P^2>0$.}
Therefore, in order to maximise the true-pion content in any measurement, kinematic configurations are chosen in order to minimise $|-t|$.  This is necessary but not sufficient to ensure the data obtained thereby are representative of the physical pion.  Additional procedures are needed in order to suppress non-resonant (non-pion) background contributions; and modern experiments and proposals make excellent use of, \emph{e.g}.\ longitudinal-transverse cross-section separation and low-momentum tagging of the outgoing nucleon.

Notwithstanding their ingenuity, such experimental techniques cannot directly address the following question: supposing it is sensible to speak of an off-shell pion with total-momentum $P$, where $P^2=(\mathpzc{v}-1)m_\pi^2$, $m_\pi \approx 0.14\,$GeV, so that $\mathpzc{v}\geq 0$ defines the pion's virtuality,
then how do the qualities of this system depend on $\mathpzc{v}$?  If the sensitivity is weak, then $\pi^\ast(\mathpzc{v})$ is a good surrogate for the physical pion; but if the distributions of, \emph{e.g}.\ charge or partons, change significantly with $\mathpzc{v}$, then the processes in Figs.\,\ref{figSullivan} can reveal little about the physical pion.  Instead, they express features of the entire compound reaction.  Since there is no unique definition of an off-shell bound-state, the question we have posed does not have a precise answer.  However, as will become clear, that does not mean there is no rational response.

\smallskip

\noindent\textbf{2.$\;$Pions: on- and off-shell}.
All correlations with pion-like quantum numbers, both resonant and continuum, are accessible via the inhomogeneous pseudoscalar Bethe-Salpeter equation:
\begin{equation}
\Gamma_{5}(k;P) = Z_4 \gamma_5 +
\int_{dq}^\Lambda [ \chi_5(q;P)  ]_{sr} K_{tu}^{rs}(q,k;P),
\label{pionBSEgeneral}
\end{equation}
where $\chi_5(q;P) =S(q_\eta) \Gamma_{5}(q;P) S(q_{\bar\eta}) $, $q_\eta =  q + \eta P$, $q_{\bar \eta} = q - (1-\eta) P$, $P$ is the total quark-antiquark momentum; $\int_{dq}^\Lambda$ represents a Poincar\'e invariant regularisation of the four-dimensional integral, with $\Lambda$ the regularisation mass-scale; and $Z_{4}(\zeta^2,\Lambda^2)$ is the mass renormalisation constant, with $\zeta$ the renormalisation point.  In addition, $S$ is the dressed-propagator for a $u$- or $d$-quark (we assume isospin symmetry throughout), $K$ is the quark-antiquark scattering kernel, and the indices $r,s,t,u$ denote the matrix structure of the elements in the equation.

The physical ($\mathpzc{v}=0$) pion appears as a pole in the pseudoscalar vertex, \emph{viz}.\ \cite{Maris:1997hd}
\begin{equation}
\label{rhopi}
\Gamma_{5}(k;P) \stackrel{P^2+m_\pi^2 \simeq 0}{=}
\frac{ \rho_\pi^\zeta}{P^2+m_\pi^2} \Gamma_\pi(k;P) + \mbox{reg.,}
\end{equation}
where ``reg.'' denotes terms analytic on $\mathpzc{v}m_\pi^2 \simeq 0$,
\begin{align}
\nonumber \Gamma_\pi&(k;P)   =   \gamma_5 \left[ i E_\pi(k;P) +
\gamma\cdot P F_\pi(k;P) \right. \\
&
\left. + \gamma\cdot k \,k \cdot P\, G_\pi(k;P)
+ \sigma_{\mu\nu}\,k_\mu P_\nu \,H_\pi(k;P)
\right]
\label{pionBSA}
\end{align}
is the pion's Bethe-Salpeter amplitude and $\rho_\pi^\zeta$ measures the ratio of the in-pion condensate and the pion's leptonic decay constant \cite{Brodsky:2012ku}.

In proposing reactions like those in Fig.\,\ref{figSullivan} as paths to real-pion targets, one is na\"{\i}vely thought to assume that for some nonzero and sizeable $\mathpzc{v}_S$, the pion pole remains the dominant feature of the pseudoscalar vertex and the pion's wave function is ``frozen'':
\begin{equation}
\label{ApproxOne}
\Gamma_{5}(k;P) \stackrel{\mathpzc{v}<\mathpzc{v}_S}{\approx}
\frac{ \rho_\pi^\zeta}{P^2+m_\pi^2} \Gamma_\pi(k;P).
\end{equation}
With modern methods of experiment and analysis, however, the reactions in Figs.\,\ref{figSullivan} provide sound realisations of a pion target under softer assumptions; namely,
the pole associated with the ground-state pion remains the dominant feature of the vertex (equivalently, the quark-antiquark scattering matrix) and the Bethe-Salpeter-like amplitude describing the related correlation evolves slowly and smoothly with virtuality.  Under these conditions, then $\forall \mathpzc{v}< \mathpzc{v}_S$ a judicious extrapolation of a cross-section to $\mathpzc{v}=0$ will yield a valid estimate of the desired on-shell result.
The question posed in the Introduction may now be translated into the challenge of determining the value of $\mathpzc{v}_S$ for which these conditions are satisfied.

To address this issue, we consider the following modified Bethe-Salpeter equation \cite{Frank:1995uk}:
\begin{equation}
\Gamma_{5}(k;P) = Z_4 \gamma_5
 + \lambda(\mathpzc{v}) \int_{dq}^\Lambda [ \chi_{5}(q;P) ]_{sr} K_{tu}^{rs}(q,k;P)\,,
\label{pionBSEoffshell}
\end{equation}
because the quantity $\delta(\mathpzc{v}):=[\lambda(\mathpzc{v})-1]$ can rigorously be said to measure deviations induced by nonzero pion virtuality.
Namely, given any value of $P^2=(\mathpzc{v}-1) m_\pi^2$, there is a unique value $\lambda(\mathpzc{v})$ for which Eq.\,\eqref{pionBSEoffshell} exhibits an (off-shell) pion pole at $(\mathpzc{v}-1) m_\pi^2$.
Subsequently, a comparison between the Bethe-Salpeter amplitude obtained at that pole and the $\mathpzc{v}=0$ amplitude will reveal the nature of (any) changes in the internal structure of the associated correlation.\footnote{
Off-shell mesons are typically defined more simply \cite{Hatsuda:1993sq, OConnell:1995nse, Benayoun:2007cu, Mitchell:1996dn, Tandy:1997qf, Tandy:1998cg, ElBennich:2011py, El-Bennich:2016bno}.  For example, in Refs.\,\cite{Mitchell:1996dn, Tandy:1997qf, Tandy:1998cg, ElBennich:2011py, El-Bennich:2016bno} the internal structure is assumed to be frozen and off-shell features, when incorporated, are expressed solely through the virtuality dependence of a vacuum polarisation diagram built using the frozen amplitudes.
}
The value of $\mathpzc{v}_S$ is the boundary of the $\mathpzc{v}$-domain for which any such modifications are modest.
(Here, ``modest'' means that all quantitative measures of structural change evolve slowly and smoothly with $\mathpzc{v}$.)
Notably, since the equation describing the pole's residue, \emph{i.e}.\ the related homogeneous Bethe-Salpeter equation, is the same in any channel that possesses overlap with the pion, then for the purpose of elucidating the character of an off-shell pion, it suffices completely to consider Eq.\,\eqref{pionBSEoffshell}.

\smallskip

\noindent\textbf{3.$\;$Computed properties of an off-shell pion}.
Hitherto, there are neither ambiguities nor model assumptions; and the character of an off-shell pion can be assessed by any nonperturbative approach that provides access to the solution of Eq.\,\eqref{pionBSEoffshell}.  We choose to approach the problem using methods developed for the continuum bound-state problem \cite{Maris:2003vk, Roberts:2015lja, Horn:2016rip, Eichmann:2016yit}.

The kernel of Eq.\,\eqref{pionBSEoffshell} involves the dressed light-quark propagators, so it is coupled with the light-quark gap equation.   The problem can therefore be analysed by using a symmetry-preserving truncation of this pair of equations.  A systematic scheme is described in Refs.\,\cite{Munczek:1994zz, Bender:1996bb, Binosi:2016rxz}; and the leading-order term is the widely-used rainbow-ladder (RL) truncation.  It is known to be capable of delivering a good description of $\pi$- and $K$-mesons \cite{Maris:2003vk, Roberts:2015lja, Horn:2016rip, Eichmann:2016yit}, for example, because corrections in these channels largely cancel owing to the preservation of relevant Ward-Green-Takahashi identities.

A more realistic description is provided by the class of symmetry-preserving DB kernels \cite{Chang:2009zb}, \emph{i.e}.\ dynamical chiral symmetry breaking (DcsB) improved kernels, which shrink the gap between nonperturbative continuum-QCD and the \emph{ab initio} prediction of bound-state properties \cite{Binosi:2014aea, Binosi:2015xqk, Binosi:2016nme}.  A basic difference between the two is that DB kernels produce a smoother transition between the weak- and strong-coupling domains of QCD, something that is expressed in mesons, \emph{e.g}.\ via softer leading-twist parton distribution amplitudes (PDAs) \cite{Chang:2013pq, Segovia:2013eca, Shi:2015esa}.  Having made the distinctions clear, we now note that the RL truncation is adequate herein because we aim to explore contrasts between bound-state properties off- and on-shell, and differences between RL and DB results will largely cancel in such ratios.

In RL truncation, the relevant gap- and Bethe-Salpeter equations are ($p=k-q$, $T_{\mu\nu}(p)=\delta_{\mu\nu}-p_\mu p_\mu/p^2$) \cite{Maris:1997tm, Bloch:2002eq, Qin:2011dd}:
\begin{subequations}
\label{gendseN}
\begin{align}
S^{-1}(k) 
& = Z_2 \,(i\gamma\cdot k + m^{\rm bm}) + \Sigma(k)\,,\\
\Sigma(k)& =  Z_2^2 \int^\Lambda_{dq}\!\! \overline{\mathpzc G}(p^2)
\,T_{\mu\nu}(p)\frac{\lambda^a}{2} \gamma_\mu S(q) \frac{\lambda^a}{2} \gamma_\nu\, ,
\end{align}
\end{subequations}
where $Z_2$ is the quark wave function renormalisation; and
\begin{align}
\nonumber
\Gamma_{5}(&k;P)  = Z_4 \gamma_5\\
& -  \lambda(\mathpzc{v})  Z_2^2 \! \int_{dq}^\Lambda
\overline{\mathpzc G}(p^2) \,T_{\mu\nu}(p) \frac{\lambda^a}{2}\,\gamma_\mu \chi_{5}(q;P)  \frac{\lambda^a}{2}\,\gamma_\nu\,.
\label{RLbse}
\end{align}

Eqs.\,\eqref{gendseN}, \eqref{RLbse} are complete once the process-in\-de\-pen\-dent running interaction is specified; and we use \cite{Qin:2011dd, Qin:2011xq}
\begin{equation}
\overline{\mathpzc G}(s) =\frac{8 \pi^2}{\omega^5} \varsigma^3 \, {\rm e}^{-s/\omega^2}
+ \frac{8 \pi^2 \gamma_m\,{\cal F}(s)}{\ln [ \tau + (1+s/\Lambda_{\rm QCD}^2)^2]} ,
\label{mathpzcG}
\end{equation}
where $\gamma_m = 12/25$, $\Lambda_{\rm QCD}=0.234\,$GeV; $\tau={\rm e}^2-1$\; ${\cal F}(s) = \{1 - \exp(-s/[4 m_t^2])\}/s$, $m_t=0.5\,$GeV; $\varsigma = 0.8\,$GeV, $\omega=m_t$; and a renormalisation scale $\zeta=\zeta_{19}=19\,$GeV \cite{Maris:1997tm}.
The connection between Eq.\,\eqref{mathpzcG} and QCD's gauge sector is canvassed elsewhere \cite{Binosi:2014aea, Binosi:2015xqk, Binosi:2016nme}.  Here we only note that Eq.\,\eqref{mathpzcG} has the correct shape but is too large in the infrared, for reasons that are well understood.  Notwithstanding this, used judiciously in RL truncation, Eq.\,\eqref{mathpzcG} serves as a valuable tool for hadron physics phenomenology.  (Notably, for a wide range of observables, Eq.\,\eqref{mathpzcG} produces results that are practically equivalent to those computed using earlier parametrisations \cite{Maris:1997tm, Maris:1999nt}.)

Solving Eq.\,\eqref{gendseN} for the dressed propagator, $S(k)=1/[i\gamma\cdot k A(k^2) + B(k^2)]$, is now straightforward; and, with the solution in hand, the kernel of Eq.\,\eqref{RLbse} is fully determined.  Thus, using $m^{\zeta_{19}}=3.4\,$MeV, at the on-shell point, $\lambda(\mathpzc{v}=0)=1$, we obtain \cite{Qin:2011xq}: $m_\pi= 0.134\,$GeV, $f_\pi=0.093\,$GeV in fair agreement with experiment \cite{Olive:2016xmw}.

\begin{figure}[t]
\begin{center}
\includegraphics[width=0.8\linewidth]{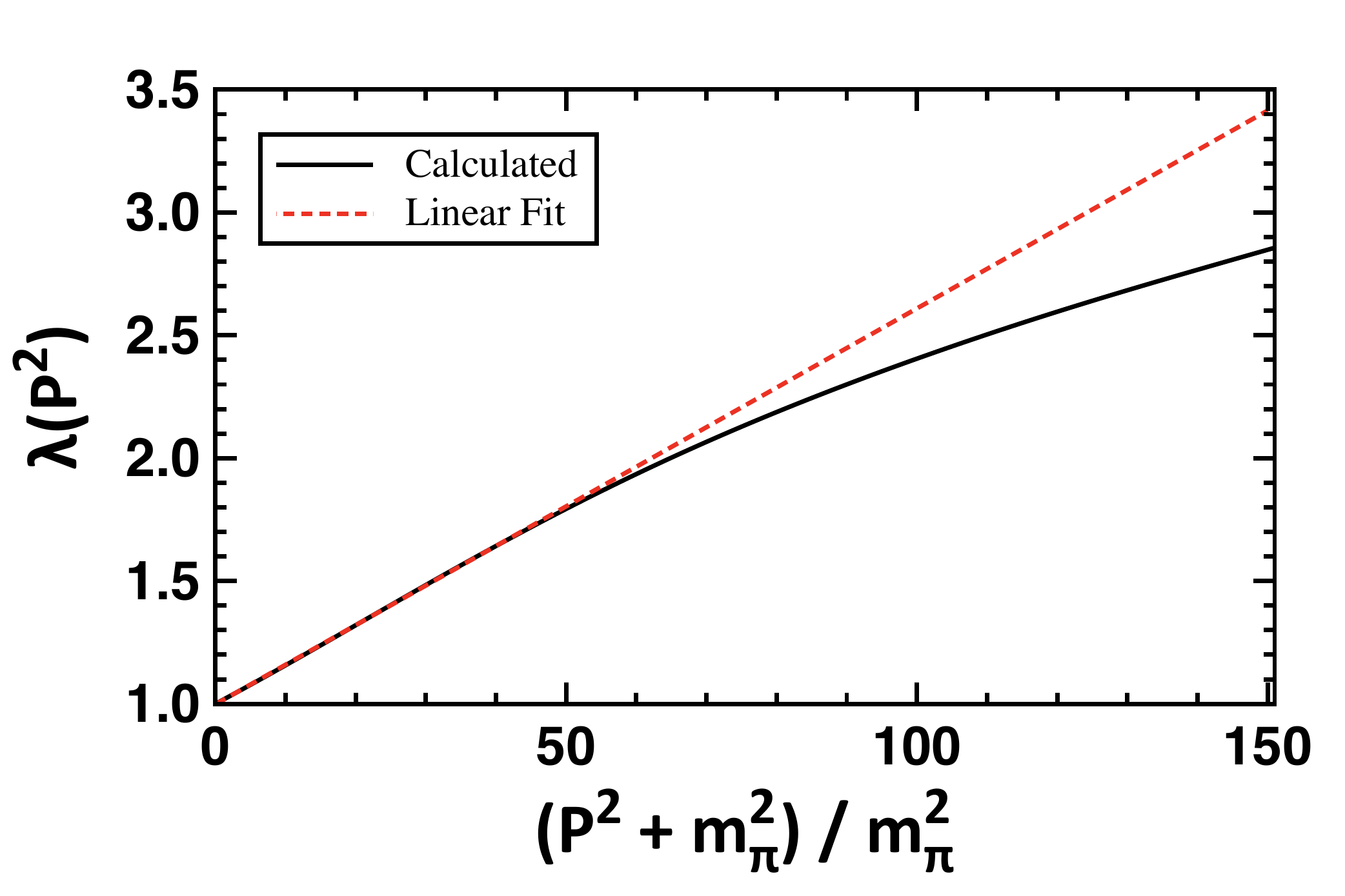}
\includegraphics[width=0.8\linewidth]{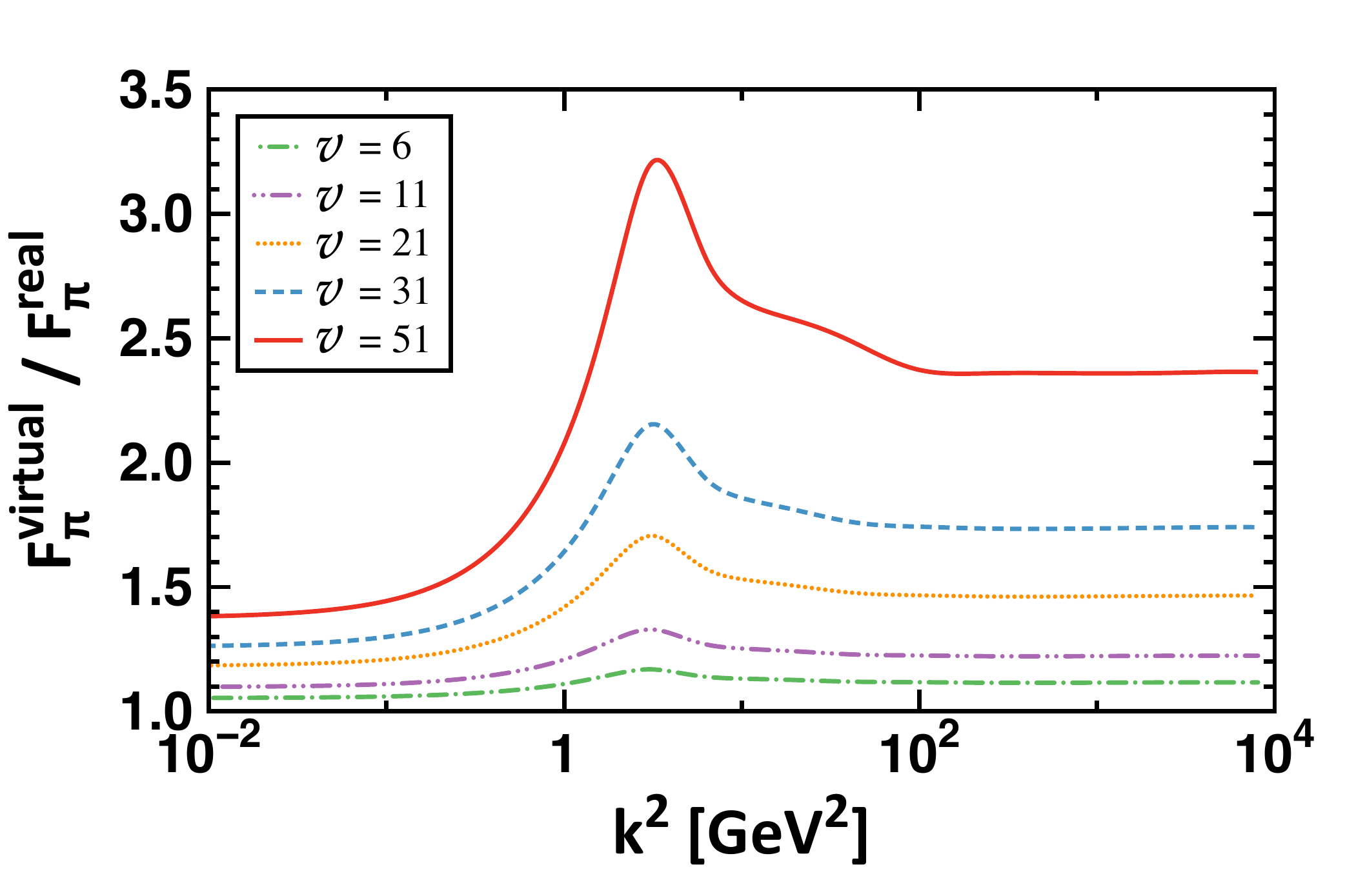}
\end{center}
\caption{\emph{Upper panel}.  $\mathpzc{v}$-dependence of the virtuality eigenvalue introduced in Eq.\,\eqref{pionBSEoffshell}.  The curve is linear on $\mathpzc{v}\lesssim 45$, Eq.\,\eqref{eqlambdav}, a result which indicates that the pion pole dominates the quark-antiquark scattering kernel on this domain.
\emph{Lower panel}.  $\mathpzc{v}$-dependence exhibited by one of the UV-dominant terms in the pion's Bethe-Salpeter amplitude, Eq.\,\eqref{pionBSA}.
\label{figlambda}
\label{figEFratio}
}
\end{figure}

With this foundation, we can begin to explore the persistence of pionic characteristics as one takes the correlation off-shell.  To that end, in Fig.\,\ref{figlambda} (upper panel) we depict the $\mathpzc{v}$-dependence of the virtuality eigenvalue: the result is linear on $\mathpzc{v}\lesssim 45$,
\begin{equation}
\label{eqlambdav}
\lambda(\mathpzc{v}) = 1 + 0.016\,\mathpzc{v}\,,
\end{equation}
\emph{i.e}.\ the change in $\lambda(\mathpzc{v})$ is purely kinematic and, hence, the pion pole dominates the quark-antiquark scattering kernel $\forall \mathpzc{v} < 45$.

The next issue to address is if/how the internal structure of the correlation is modified.  A detailed picture of possible rearrangements of the pion's internal structure can be obtained by studying the impact of $\mathpzc{v}> 0$ on the scalar functions in Eq.\,\eqref{pionBSA}.  This is illustrated in Fig.\,\ref{figEFratio} (lower panel), which depicts the $k^2$-dependence of the ratio of the leading Chebyshev moment for one of the ultraviolet (UV) dominant amplitudes in Eq.\,\eqref{pionBSA}, where for any function that leading moment is ($x=k\cdot P/\sqrt{k^2 P^2}$):
\begin{equation}
{\mathpzc W}(k^2;P^2) = \frac{2}{\pi} \int_{-1}^1dx\,\sqrt{1-x^2}\,{\mathpzc W}(k^2,x;P^2)\,.
\end{equation}

The evolution pattern of the correlation's internal structure is more subtle than that of $\lambda(\mathpzc{v})$.  Notwithstanding that, we find that structural modifications are significant $\forall \mathpzc{v} > 45$.  Moreover, there is a measure of ambiguity in demarcating the domain within which structural changes can be considered modest.  We therefore choose conservatively and identify $\mathpzc{v}_S \approx 31$, since on the domain $\mathpzc{v} \lesssim \mathpzc{v}_S$ the pattern exhibited by the ratios in Fig.\,\ref{figEFratio} is both simple and readily interpreted.
Namely, on $k^2\lesssim 1\,$GeV$^2$, \emph{i.e}.\ at length-scales $\ell_\pi \gtrsim 0.2\,$fm, the impact of $\mathpzc{v}\neq 0$ on the pion's internal structure is modest, even at $\mathpzc{v} = 31$.
The domain $k^2\in[1,4]\,$GeV$^2$ is a smooth region of transition into the UV.  Then,
on $k^2\gtrsim 4\,$GeV$^2$, \emph{viz}.\ for $\ell \lesssim 0.1\,$fm, one observes plateaux, which describe nearly constant shifts in the amplitudes.  The magnitude of the shifts grows with $\mathpzc{v}$ and that growth is linear to within 3.5\%.

\begin{figure}[t]
\includegraphics[width=0.8\linewidth]{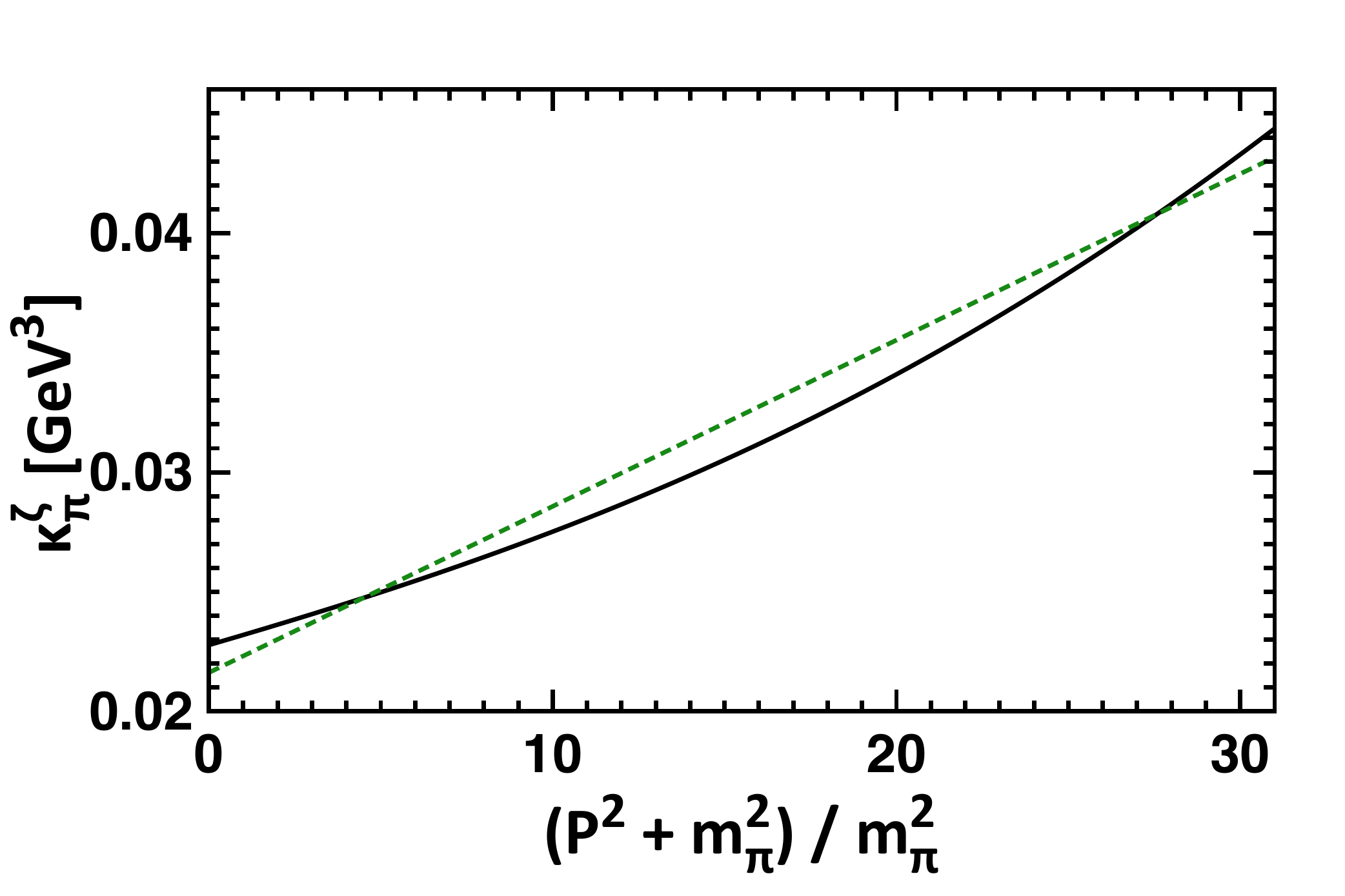}
\caption{Virtuality dependence of the quark-antiquark core density in the pion correlation: solid (black) curve.
On the depicted domain, the evolution is linear to within 3\%, as highlighted by the dashed (green) line.
(We use $\zeta=19\,$GeV.)
\label{figdensity}}
\end{figure}

The UV tail of the pion's Bethe-Salpeter amplitude maps algebraically into a $\mathpzc{v}$-dependence of $\rho_\pi^\zeta$ in Eq.\,\eqref{rhopi}:
\begin{equation}
\label{rhogen}
i\rho_{\pi}^\zeta(\mathpzc{v}) = Z_4\; {\rm tr}_{\rm CD}
\int_{dq}^\Lambda \gamma_5 \chi_{\pi}(q^2,q\cdot P;\mathpzc{v})  \,,
\end{equation}
where $\chi_\pi = S(q_\eta) \Gamma_{\pi}(q^2,q\cdot P;\mathpzc{v}) S(q_{\bar\eta})$ and the trace is over colour and spinor indices, because the value of the integral in Eq.\,\eqref{rhogen} is determined by the ultraviolet behaviour of the integrand \cite{Langfeld:2003ye}.  An analogous leptonic decay constant can also be defined:
\begin{equation}
\label{fpigen}
f_\pi(\mathpzc{v}) P_\mu = Z_4\; {\rm tr}_{\rm CD}
\int_{dq}^\Lambda \gamma_5 \gamma_\mu \chi_{\pi}(q^2,q\cdot P;\mathpzc{v}) \,.
\end{equation}
%
One can now form the product $\kappa^\zeta_{\pi}(\mathpzc{v}):=f_\pi(\mathpzc{v})\rho_{\pi}^\zeta(\mathpzc{v})$, which is a quark-antiquark core density for the correlation, an in-pion condensate \cite{Brodsky:2012ku}, whose growth with virtuality is depicted in Fig.\,\ref{figdensity}.
Unsurprisingly, given the preceding observations, $\kappa^\zeta_{\pi}(\mathpzc{v})$ grows approximately linearly with virtuality on $\mathpzc{v}\lesssim \mathpzc{v}_S$:
\begin{equation}
\kappa^\zeta_{\pi}(\mathpzc{v}) \approx \kappa^\zeta_{\pi}(0) [ 1+ 0.032 \mathpzc{v}]\,,
\;  \kappa^\zeta_{\pi}(0) =(0.28\,{\rm GeV})^3.
\end{equation}

The picture that emerges, therefore, is an off-shell pion whose internal structure is essentially unaltered at length-scales $\ell_\pi \gtrsim 0.1\,$fm.  On the other hand, at the core ($\ell_\pi \lesssim 0.1$\,fm) the quark-antiquark density increases slowly with virtuality, reaching a value at $\mathpzc{v}=31$ which is roughly twice that of the on-shell pion, in line with expectations based upon the plateaux in Fig.\,\ref{figEFratio}.  (A linear fit to $\kappa^\zeta_{\pi}(\mathpzc{v})$ on $\mathpzc{v}\in[0,55]$ is a poor representation of the result: the rms-difference is greater than 10\% and it underestimates $\kappa^\zeta_{\pi}(0)$ by 40\%.)

\begin{figure}[t]
\begin{center}
\includegraphics[width=0.85\linewidth]{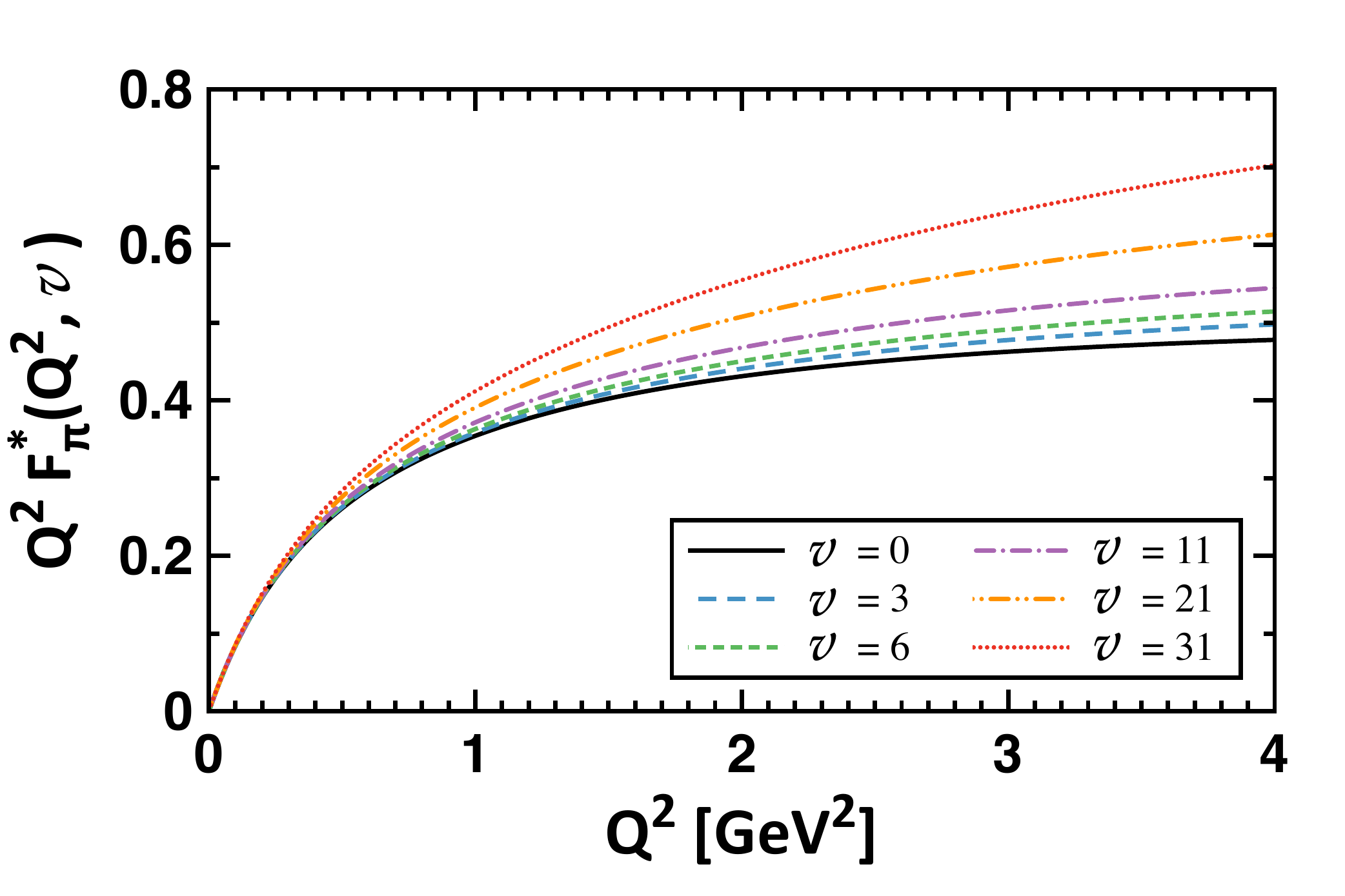}
\includegraphics[width=0.85\linewidth]{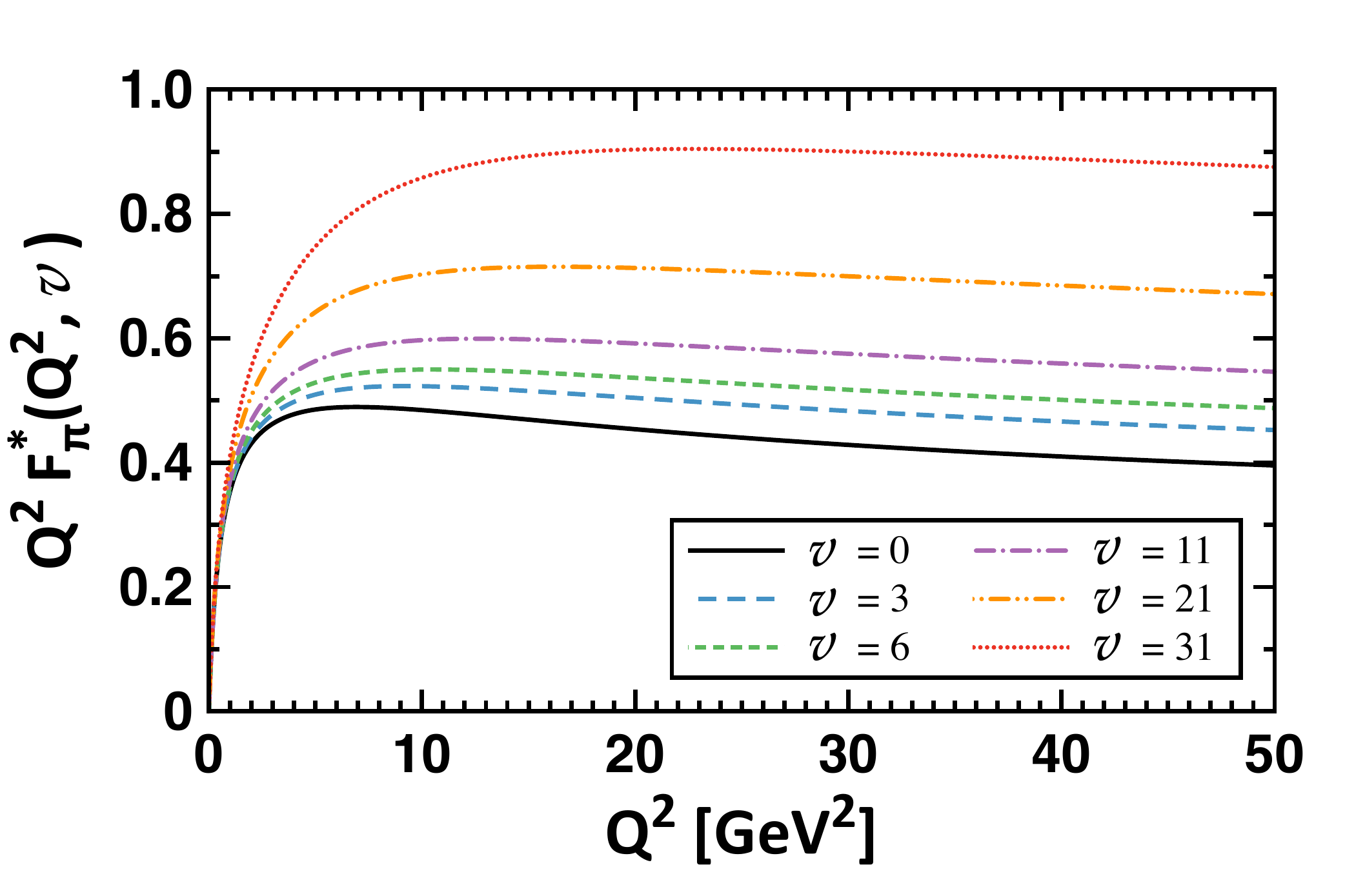}
\end{center}
\caption{\emph{Upper panel}.  Direct calculation of the $\pi^\ast(\mathpzc{v})+\gamma \to \pi$ transition form factor at a range of virtuality values.
\emph{Lower panel}.  Constrained extrapolations to large $Q^2$ using Eq.\,\eqref{FastFit}.
\label{fighalfpionFF}}
\end{figure}

As evident in Fig.\,\ref{figSullivan}, only one pion is off-shell when using the Sullivan process to generate a hadron target.  Consequently, the modest structural changes described above enter linearly in the scattering amplitudes.  Their impact is illustrated in Fig.\,\ref{fighalfpionFF}, which depicts the $\pi^\ast(\mathpzc{v})+\gamma \to \pi$ transition form factor, $F^\ast_\pi(Q^2,\mathpzc{v})$.
Using the ``brute force'' algorithm employed in Ref.\,\cite{Maris:2000sk} (to compute the propagators, Bethe-Salpeter amplitudes, photon-quark vertex, and scattering amplitude) yields the curves drawn in the upper panel of Fig.\,\ref{fighalfpionFF}.  Those curves terminate at $Q^2 = 4\,$GeV$^2$ because the algorithm is unreliable at larger momenta.

To complete the calculation of $F^\ast_\pi(Q^2,\mathpzc{v})$ directly at arbitrarily large spacelike $Q^2$, it would be necessary to use the method introduced in Ref.\,\cite{Chang:2013nia}, \emph{i.e}.\ develop a new perturbation theory integral representation for the Bethe-Salpeter amplitude at each required value of the virtuality.  That is straightforward but time consuming, so we employ a simpler expedient.  Namely, we capitalise on the analysis in Ref.\,\cite{Chang:2013nia}, which shows that the computed elastic pion form factor can accurately be interpolated by a monopole multiplied by a simple factor that restores the correct QCD anomalous dimension.  We therefore write
{\allowdisplaybreaks
\begin{subequations}
\label{FastFit}
\begin{align}
F^\ast_\pi(Q^2,\mathpzc{v}) & = \frac{1}{1+Q^2/m_0^2} \mathpzc{A}(Q^2,\mathpzc{v})\\
\mathpzc{A}(Q^2,\mathpzc{v}) & =
\frac{1 + Q^2 a_0^2(\mathpzc{v})}
{1 + Q^2 [a_0^2(\mathpzc{v}) / b_u^2(\mathpzc{v})] \ln(1+Q^2/\Lambda_{\rm QCD}^2)}
\end{align}
\end{subequations}}
\hspace*{-0.5\parindent}where $m_0= 0.72\,$GeV (\emph{i.e}., the $\rho$-meson mass computed using this framework \cite{Qin:2011dd}) is fixed by the elastic pion form factor, and $a_0(\mathpzc{v})$, $b_u(\mathpzc{v})$ are fitted to the behaviour of $F^\ast_\pi(Q^2,\mathpzc{v})$ on $Q^2\in [0,4]\,$GeV$^2$:
\begin{subequations}
\label{a0bu}
\begin{align}
a_0(\mathpzc{v}) & = 0.29 (1+0.028\,\mathpzc{v})\,, \\
b_u(\mathpzc{v}) & = 2.3 (1+0.017 \,\mathpzc{v}) \,.
\end{align}
\end{subequations}
The lower panel depicts a collection of such constrained extrapolations.  Pointwise comparison with Fig.\,2 in Ref.\,\cite{Chang:2013nia} demonstrates the veracity of Eq.\,\eqref{FastFit} for $\mathpzc{v}=0$.

An important feature of the transition form factor is highlighted by the lower panel of Fig.\,\ref{fighalfpionFF}, \emph{viz}.\ once again, on $\mathpzc{v}\lesssim \mathpzc{v}_S$ the magnitude of $F^\ast_\pi(Q^2,\mathpzc{v})$ for $Q^2\gtrsim 10\,$GeV$^2$ grows approximately linearly with $\mathpzc{v}$.  This, too, can be traced to the behaviour illustrated in Fig.\,\ref{figEFratio} (lower panel) because, reviewing the analysis in Ref.\,\cite{Maris:1998hc}, it is readily established that the UV behaviour of the $\pi^\ast(\mathpzc{v})+\gamma \to \pi$ transition form factor must respond linearly to changes in the Bethe-Salpeter amplitude and such modifications should become evident on just this domain.

\begin{figure}[t]
\begin{center}
\includegraphics[width=0.8\linewidth]{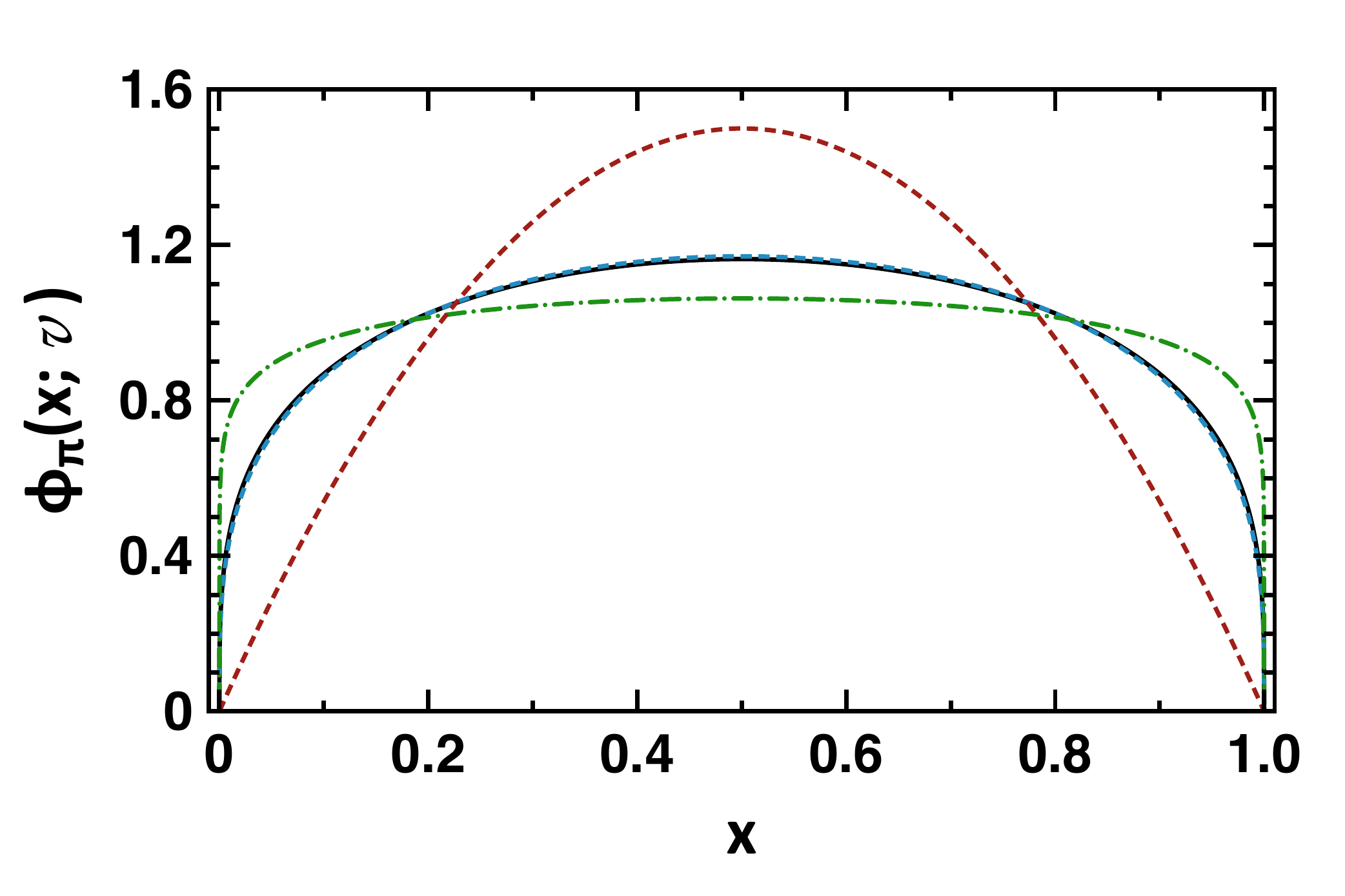}
\end{center}
\caption{Virtuality-dependence of pion twist-two PDA.  Solid curve: inferred $\mathpzc{v}=0$ result, a good approximation to that calculated in RL truncation, dashed (blue) \cite{Chang:2013pq, Chang:2013nia}; and dot-dashed (green) curve, inferred PDA at $\mathpzc{v}=31$.
Even this appreciable virtuality only introduces a modest rms relative-difference between the PDAs determined herein; namely, 13\%.  Measured equivalently, the RL result differs by 34\% from that appropriate to QCD's conformal limit (dotted, red).
\label{PDAv}}
\end{figure}

One can elaborate by recalling \cite{Lepage:1979zb, Lepage:1980fj, Efremov:1979qk}:
\begin{subequations}
\label{pionUV}
\begin{align}
Q^2 F_\pi(Q^2) & \stackrel{Q^2 \gg \Lambda_{\rm QCD}^2}{\approx} 16 \pi \alpha_s(Q^2)  f_\pi^2 \mathpzc{w}_\varphi^2, \\
\mathpzc{w}_\varphi & = \tfrac{1}{3} \int_0^1 dx\, \tfrac{1}{x} \varphi_\pi(x)\,,
\end{align}
\end{subequations}
where $\varphi_\pi(x)$ is the pion's twist-two valence-quark PDA.
Contemporary analyses demonstrate that ground-state meson PDAs are well represented by \cite{Chang:2013pq, Segovia:2013eca, Shi:2015esa, Zhang:2017bzy} $\varphi(x) = \mathpzc{N}_\mathpzc{p} [x(1-x)]^\mathpzc{p}$, where $\mathpzc{N}_\mathpzc{p}$ ensures $\int_0^1 dx \varphi(x) =1$.
Moreover, when the consistently-computed PDA is used, Eq.\,\eqref{pionUV} underestimates the direct RL calculation by only 15\% on $Q^2 \simeq 8\,$GeV$^2$.  One may therefore equate Eq.\,\eqref{pionUV} with 85\% of the UV limit of Eq.\,\eqref{FastFit} and infer $\mathpzc{p}$.  This procedure yields $\mathpzc{p}(\mathpzc{v}=0)=0.29$, to be compared with $\mathpzc{p}=0.30$ in Ref.\,\cite{Chang:2013nia}, thereby confirming its validity and also the remark following Eqs.\,\eqref{a0bu}.\footnote{Direct comparison is meaningful because Ref.\,\cite{Chang:2013nia} neglected evolution of the pion's Bethe-Salpeter wave function, whose role and importance is discussed in Refs.\,\cite{Raya:2015gva, Raya:2016yuj}.}
For $\mathpzc{v}>0$, Eq.\,\eqref{pionUV} receives minor modifications: $f_\pi^2 \to f_\pi f_\pi(\mathpzc{v})$ and $\mathpzc{w}_\varphi^2\to \mathpzc{w}_\varphi \mathpzc{w}_{\varphi(\mathpzc{v})}$, where $\varphi(x;\mathpzc{v})$ is a PDA for the off-shell pion.
Using the revised formula in the matching procedure and assuming the offset remains at 15\%, then $\mathpzc{p}(\mathpzc{v}=31)=0.105$.
%
%
This inferred virtuality-dependence of the PDA is depicted in Fig.\,\ref{PDAv}: the dilation grows modestly with increasing $\mathpzc{v}$.  Such a connection between the UV behaviour of the pion's Bethe-Salpeter amplitude and dilation of the PDA is readily verified using a simple generalisation of the algebraic model introduced in Ref.\,\cite{Chang:2013pq}.

At this point, we use generalised parton distributions (GPDs) to translate the behaviour of $F^\ast_\pi(Q^2,\mathpzc{v})$ into insights regarding the impact of virtuality on extractions of the pion's valence-quark PDF via the process in Fig.\,\ref{figSullivan}(b).  In particular, recall that the elastic form factor can be written \cite{Ji:1996nm, Radyushkin:1997ki, Mueller:1998fv}:
{\allowdisplaybreaks
\begin{subequations}
\label{GPD1}
\begin{align}
F_\pi(Q^2) & = \int_{-1}^{1} dx \, H^u_{\pi^+}(x,0,Q^2)\,,\\
u^\pi(x) & = H_{\pi^+}^u(x>0,0,0) \,,
\end{align}
\end{subequations}}
\hspace*{-0.5\parindent}where $H^u_{\pi^+}(x,0,Q^2)$ is the pion's GPD and $u^\pi(x)$ is its valence-quark distribution function.  Notably, too, at a typical hadronic scale \cite{Yuan:2003fs}:
\begin{equation}
\label{GPD2}
H^u_{\pi^+}(x,0,Q^2) \stackrel{x\simeq 1}{\sim} (1-x)^2 \quad \forall Q^2 < \infty\,.
\end{equation}
Hence, considering a half off-shell generalisation of the GPD, which may be accomplished following Ref.\,\cite{Mezrag:2014jka}, using a matrix element defined with an initial state corresponding to the lowest-mass pole solution of Eq.\,\eqref{pionBSEoffshell}, and given the modest $\mathpzc{v}$-dependence of $F^\ast_\pi(Q^2,\mathpzc{v})$, Eqs.\,\eqref{GPD1}, \eqref{GPD2} indicate that $u^\pi(x;\mathpzc{v})$ will behave similarly.  In particular, the power-law describing its decay on $x\simeq 1$ should not depend strongly on $\mathpzc{v}$.

\smallskip

\noindent\textbf{4.$\;$Conclusion}.
One can define and explore the properties of an off-shell pion by introducing a virtuality eigenvalue, $\lambda(\mathpzc{v})$, into the Bethe-Salpeter equations describing the formation of bound-states and correlations in scattering channels that overlap with the pion.  The pion pole dominates the scattering matrix so long as $\lambda(\mathpzc{v})$ is linear in the virtuality, $\mathpzc{v}$.  Within this linearity domain, alterations of the pion's internal structure induced by $\mathpzc{v}>0$ can be analysed by charting the $\mathpzc{v}$-dependence of the pointwise behaviour of the Bethe-Salpeter amplitude describing the correlation.
Following this procedure, we demonstrated that for $\mathpzc{v} \lesssim \mathpzc{v}_S = 31$,  which corresponds to $-t\lesssim 0.6\,$GeV$^2$  in the notation of Fig.\,\ref{figSullivan}, the off-shell correlation serves as a valid pion target. Namely, on this domain the properties of the off-shell correlation are simply related to those of the on-shell pion and, consequently, a judicious extrapolation to $\mathpzc{v}=0$ will deliver reliable results for pion properties.

In the present context it is natural to ask for a similar statement concerning the kaon.  We have addressed this issue by repeating the analysis described herein for a fictitious $s+\bar s$ pseudoscalar bound-state.  Using a $s$-quark current-mass that produces the empirical $\phi$-meson mass \cite{Qin:2011xq}, we obtain $m_{{s\bar s}_{0^-}}=0.70\,$GeV and find $\mathpzc{v}_S^{{s\bar s}_{0^-}}= 2.7$ (units of $m_{{s\bar s}_{0^-}}^2$).  Interpolating to the kaon mass, we estimate that an off-shell correlation in this channel can serve as a valid meson target on $-t\lesssim 0.9\,$GeV$^2$.  


\smallskip

\noindent\textbf{Acknowledgments}.
We are grateful to R.~Ent, T.~Horn, C.~Keppel and J.~Rodr\'{\i}guez-Quintero for insightful comments.
Research supported by:
Argonne National Laboratory, Office of the Director, through the Named Postdoctoral Fellowship Program;
Funda\c{c}\~ao de Amparo \`a Pesquisa do Estado de S\~ao Paulo - FAPESP Grant No. 2015/21550-4;
and U.S.\ Department of Energy, Office of Science, Office of Nuclear Physics, contract no.~DE-AC02-06CH11357.


\end{document}